\documentclass{cs20proc}

\usepackage{kantlipsum}
\usepackage{url}
\editors{S. J. Wolk}
\publisher{Zenodo}
\conference{The 20th Cambridge Workshop on Cool Stars, Stellar Systems, and the Sun}
\conferencedate{2018}

\title{The Substellar Transition Zone: A Stretched Temperature Canyon in Brown Dwarf Population due to Unsteady Hydrogen Fusion}
\author{ZengHua Zhang}

\affiliation{
GEPI, Observatoire de Paris, Universit{\'e} PSL, CNRS, 5 Place Jules Janssen, 92190 Meudon, France 
}

\shorttitle{The Substellar Transition Zone}
\shortauthors{ZengHua Zhang}

\abs{I summarize results on transitional and degenerate brown dwarfs presented in a series titled {\sl Primeval very low-mass stars and brown dwarfs}. I introduce an L subdwarf classification scheme, which classified  L subdwarfs into three metal subclasses. I would also like to draw your attention to transitional brown dwarfs which have long-lasting unsteady hydrogen fusion in their cores to replenish the dissipation of their initial thermal energy. The mass range of transitional brown dwarfs with solar metallicity is between 0.065 and 0.079 m$_{\odot}$ according to the latest evolutionary models. The temperature distribution of transitional brown dwarfs are stretched to a wide range and formed a substellar transition zone. The substellar transition zone is most significant in the old halo population and ranges from 1000 K to 2200--3000 K depending on metallicity. The transition zone has impacts on our observational properties of field brown dwarf population. Because field L dwarfs are composed of very low-mass stars, transitional brown dwarfs, and relative younger electron-degenerate brown dwarfs.}

\begin{document}

\maketitle

\section{Introduction}
Brown dwarfs (BDs) are substellar objects that were predicted in the 1960s \citep{kuma63,haya63} and discovered in the 1990s \citep{rebo95,naka95}. They are the low-mass end of the initial mass function (IMF), and are very sensitive for IMF measurement. Known BDs have mass between $\sim$ 3 and 90 M$_{\rm Jup}$ and temperature ($T_{\rm eff}$) between 250 and 2800 K, which are overlapped with those of gaseous exoplanets, thus are key to understand ultra-cool atmospheres and characterize exoplanets. 

The characterization of BD population remains a big challenge. Some spectral features of BDs are not well reproduced by the latest atmospheric models \citet{alla14}. The mass of stellar/substellar boundary measured by different approaches have notable differences. First because BDs are faint, and have ultra-cool atmospheres, complex molecular absorption lines. Secondly, BDs have mass/age degeneracy, their $T_{\rm eff}$ are age dependent. It is difficult to characterize a BD without knowing its mass or age which are difficult to measure. Thirdly, the `substellar transition zone' formed by transitional BDs (T-BDs) between very low-mass stars (VLMS) and electron-degenerate BDs (D-BDs) was not considered previously in the characterization of brown dwarfs. 

In this paper I summarize those results on L subdwarfs, T-BDs, D-BDs, and the substellar transition zone presented in a series titled {\sl Primeval very low-mass stars and brown dwarfs} \citep{zha17a,zha17b,zha18a,zha18b,zha19a,zha19b}\footnote{Optical and near infrared spectra of new L subdwarfs presented in the first four papers of the series are available online: \url{https://academic.oup.com/mnras/article-lookup/doi/10.1093/mnras/sty2054#supplementary-data}}.

\begin{figure*}[ht]
	\centering
	\includegraphics[width=\linewidth]{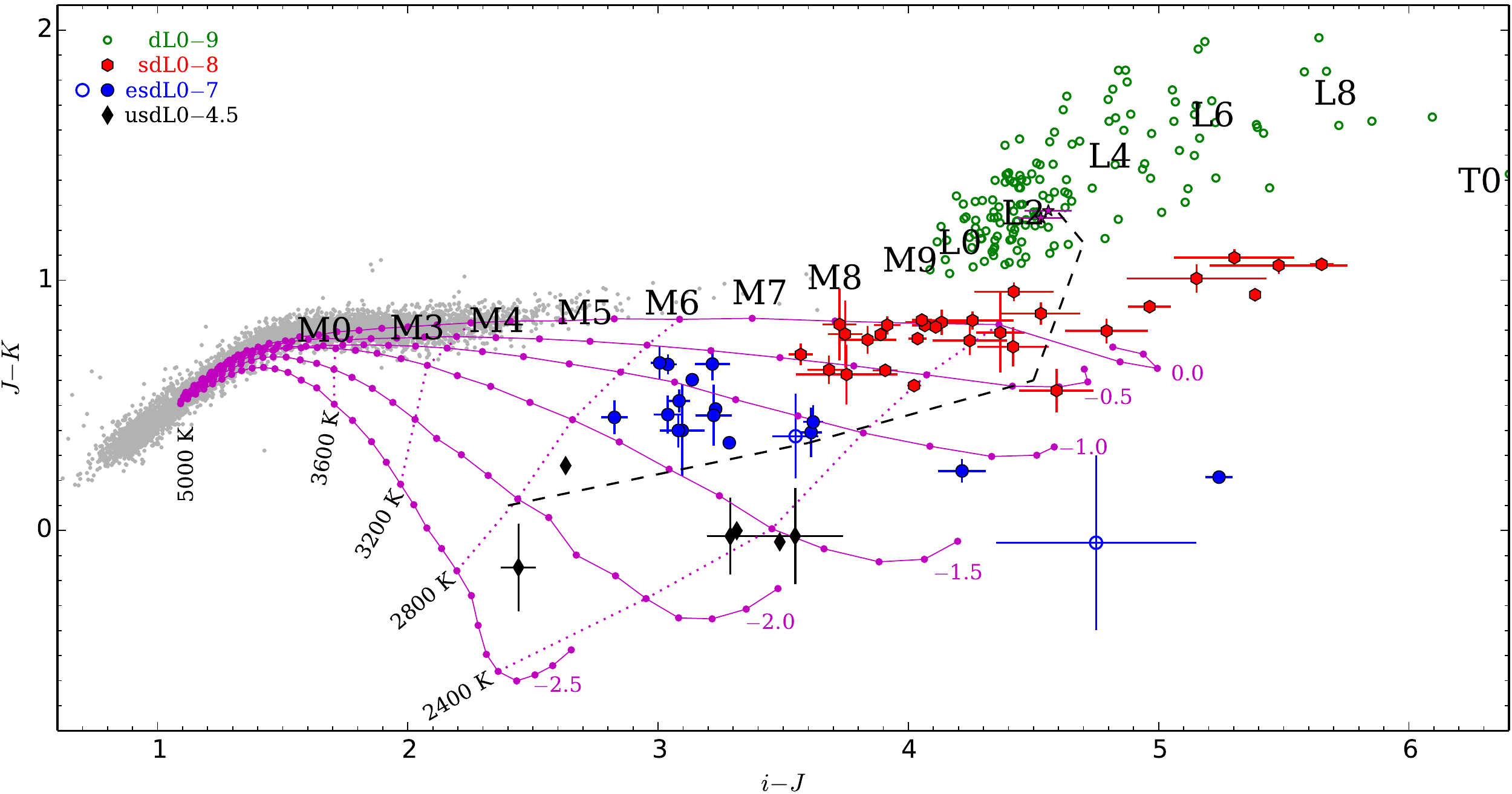}
	\caption{The $i-J$ vs $J-K$ colour-colour plot of dL, sdL, esdL and usdL subclasses compared to main-sequence and L dwarfs. A black dashed line indicates the approximate stellar/substellar boundary \citep[fig. 17 of][]{zha18b}.}
	\label{fig:ijk}
\end{figure*}

\begin{figure}
	\centering
	\includegraphics[width=\linewidth]{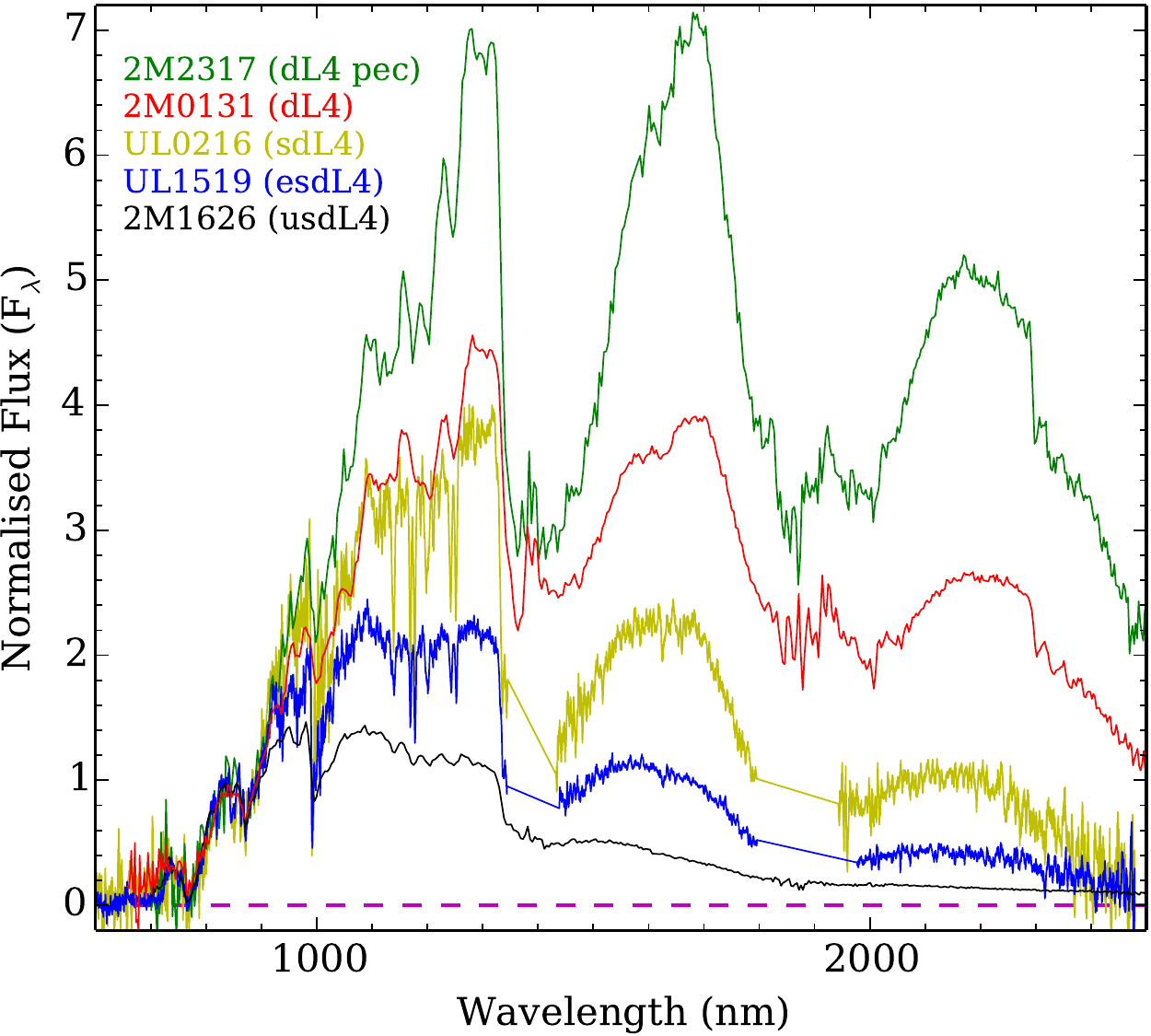}
	\caption{Spectra of red dL4, dL4, sdL4, esdL4, and usdL4 subclasses \citep[modified from fig. 9 of][]{zha17a}. Spectrum of 2MASS J23174712-4838501 (2M2317) is from \citet{kirk10}, 2MASS J01311838+3801554 (2M0131) is from \citet{burg10}, ULAS J021642.96+004005.7 (UL0216) and ULAS J151913.03$-$000030.0 are from \citet{zha17a}, 2MASS J16262034+3925190 (2M1626) is from \citet{burg04}. } 
	\label{fig:spec}
\end{figure}

\begin{figure}
	\centering
	\includegraphics[width=\linewidth]{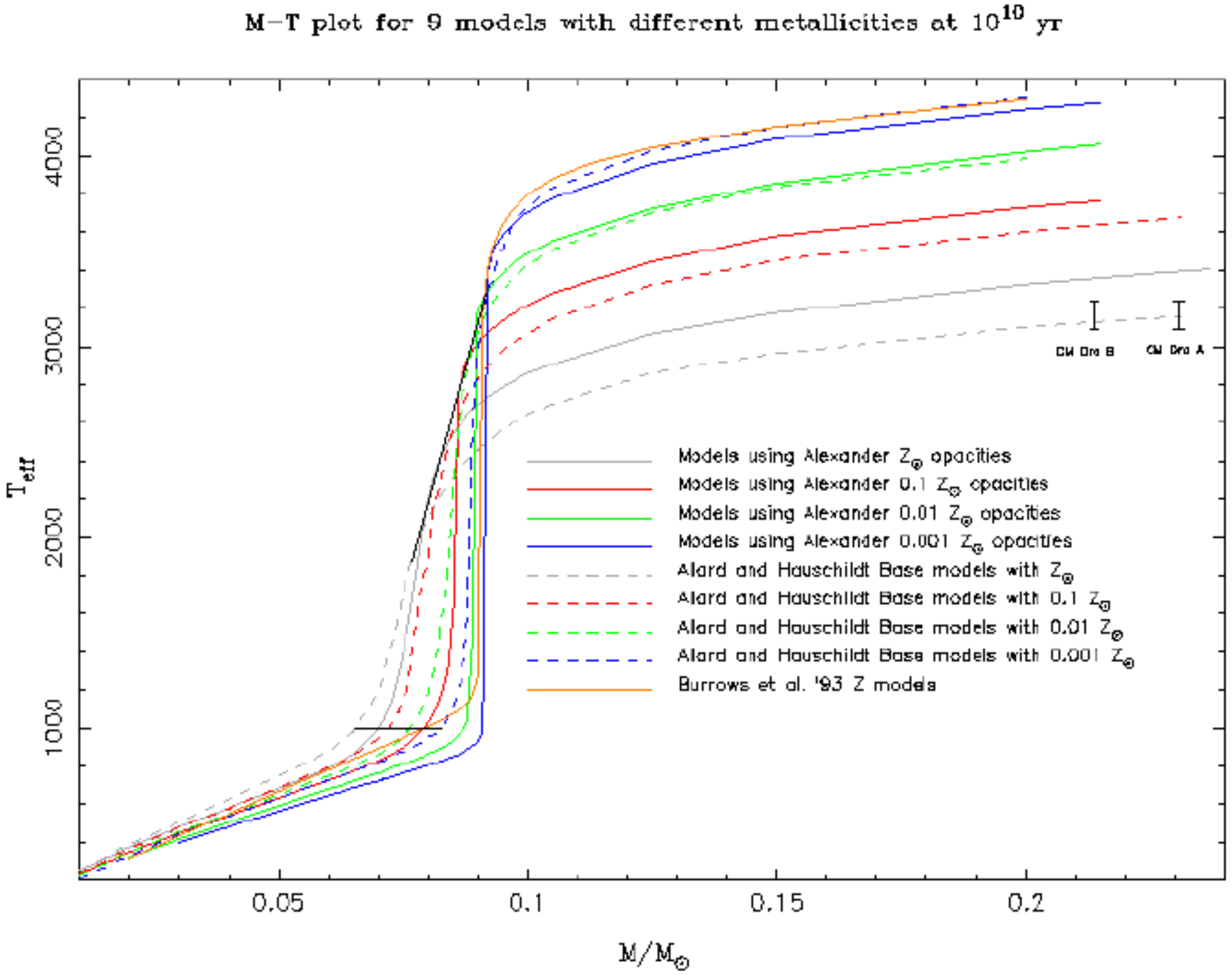}
	\caption{The mass--$T_{\rm eff}$ isochrones at 10 Gyr for low-mass objects with different metallicity \citep[modified from fig. 5 of][]{burr01}. The black line on top left indicates the SHBMM or the stellar boundary on these dashed lines \citep[based on models of][]{alla95}. A substellar transition zone is below the SHBMM and is likely Stretched to around  $T_{\rm eff}$ = 1000 K at 10 Gyr (between two black short lines). The mass width of the transition zone is around 0.012 M$_{\odot}$ at 1--0.01 Z$_{\odot}$ and around 0.0085 M$_{\odot}$ at 0.001 Z$_{\odot}$.}
	\label{fig:burr01}
\end{figure}

\begin{figure}
	\centering
	\includegraphics[width=\linewidth]{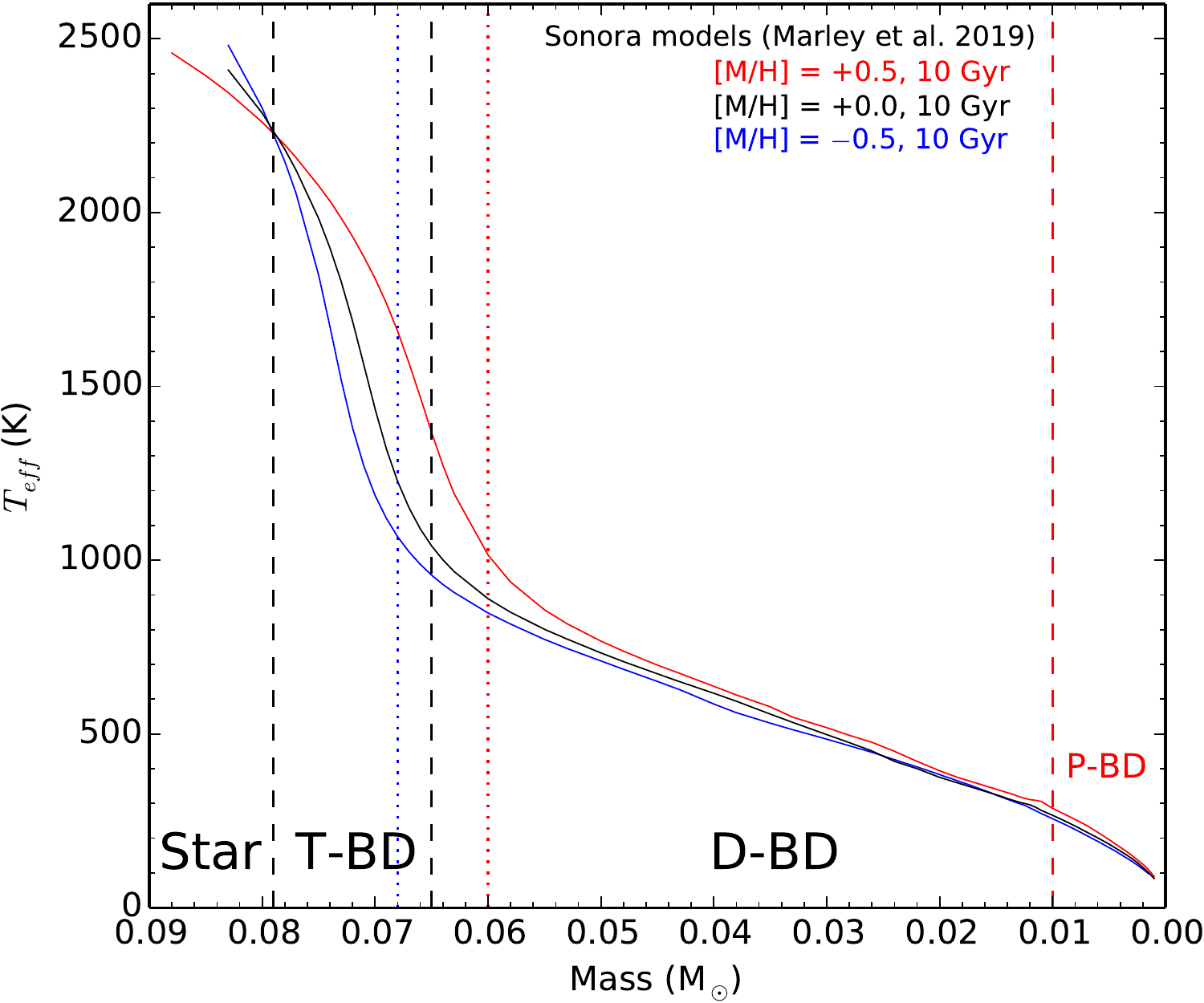}
	\caption{The 10 Gyr isochrones of $T_{\rm eff}$ of T-BDs and D-BDs with metallicities of [M/H] = +0.5, 0.0, and $-$0.5 at 10 Gyr predicted by Sonora models (Marley et al. 2019, in prop.). The intersection between [M/H] = 0.0 and [M/H] = $\pm$0.5 around 0.069 M$_{\odot}$ defines the stellar--substellar boundary at solar metallicity. The substellar transition zone of T-BDs with [M/H] = 0.0 is between these two dashed lines (also see Fig. \ref{fig:gr}). 
}
	\label{fig:teff}
\end{figure}

\begin{figure}
	\centering
	\includegraphics[width=\linewidth]{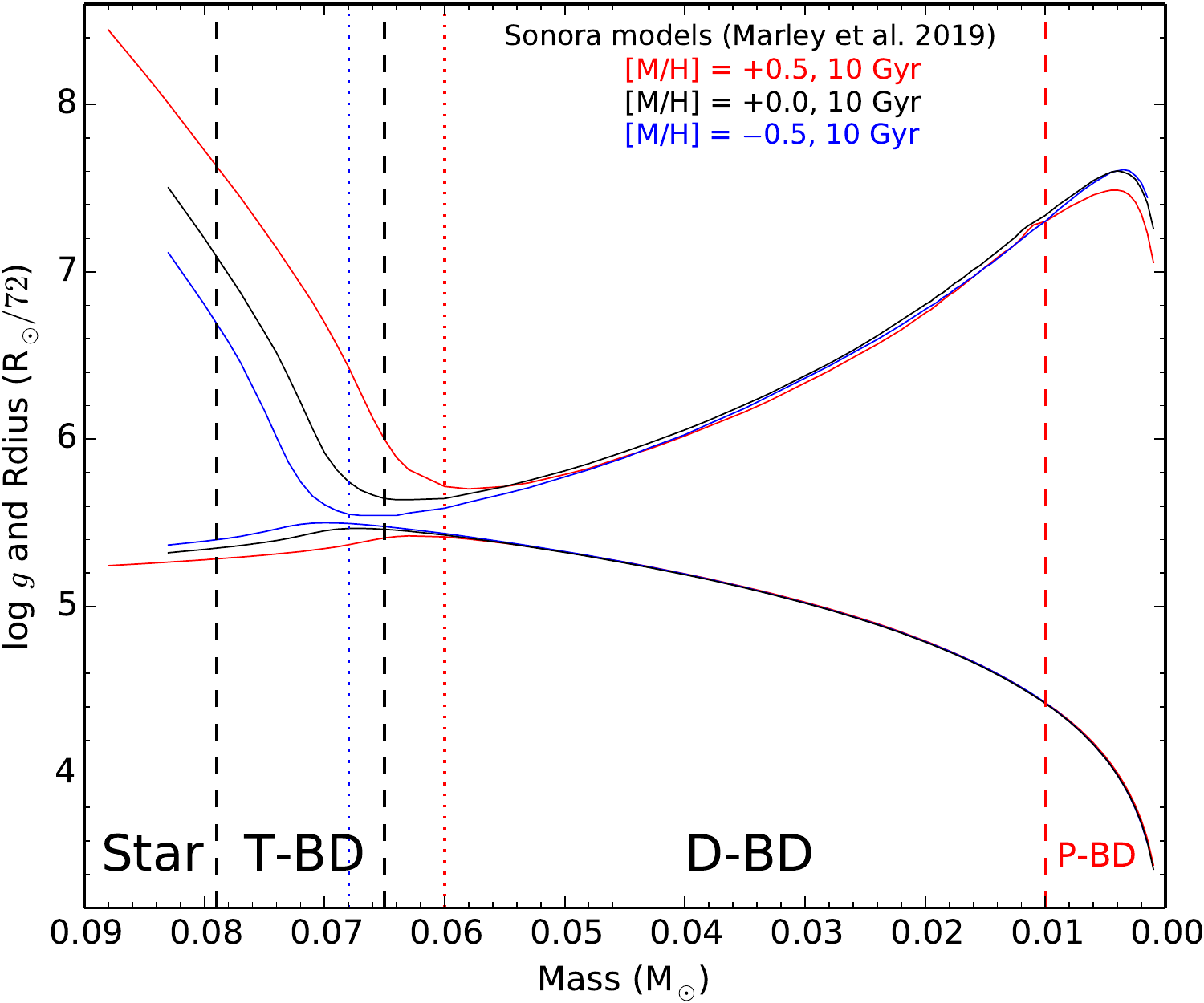}
	\caption{The 10 Gyr isochrones of gravity (lower) and radius (upper) of T-BDs and D-BDs with metallicities of [M/H] = +0.5, 0.0, and $-$0.5 at 10 Gyr predicted by Sonora models (Marley et al. 2019, in prop.). Note the unit of radius isochrones is scaled to R$_{\odot}/72$ for better comparisons with gravity isochrones. The mass with maximum gravity and minimum radius define the boundary between T-BDs and D-BDs, which is at 0.068 M$_{\odot}$ at [M/H] = $-0.5$, 0.065 M$_{\odot}$ at [M/H] = 0.0, and 0.06 M$_{\odot}$ at [M/H] = $+0.5$ (indicated with dotted lines). 
}
	\label{fig:gr}
\end{figure}


\begin{figure*}[ht]
	\centering
	\includegraphics[width=\linewidth]{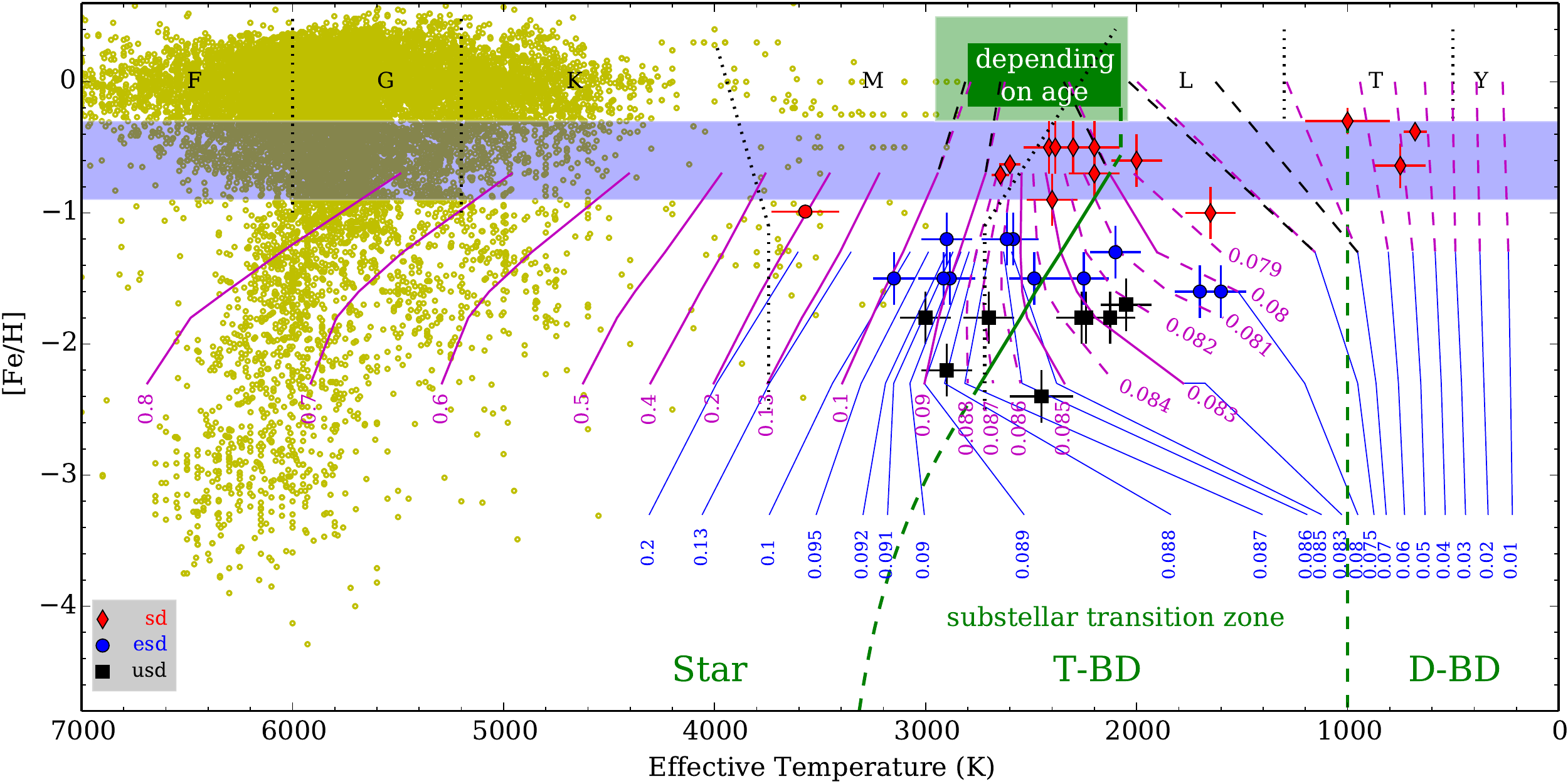}
	\caption{$T_{\rm eff}$ versus [Fe/H] of dwarf populations \citep[modified from fig. 9 in][]{zha17b}. The magenta \citep{bara97}, black \citep{bara15}, and blue \citep{burr98} solid/dashed lines indicate iso-mass contours at 10 Gyr. Objects of different spectral classes are divided by black dotted lines. Stars, transitional and degenerate brown dwarfs are divided by solid and dashed green lines. A substellar transition zone is revealed by the 10-Gyr iso-mass counter (magenta and blue lines; labelled in M$_{\odot}$) between the stellar--substellar boundary and $T_{\rm eff} \approx 1000$ K \citep[two green lines; equations 1 and 2 of][]{zha17b}. Yellow circles represent the locus of main sequence stars. This programme is targeting L subdwarfs around this stellar--substellar boundary and in the substellar transition zone.
	}
	\label{fig:tm}
\end{figure*}

\section{Classification of L subdwarfs}
L subdwarfs have rather diverse spectral features because they are distributed into a wide range of metallicity (and opacity), and have cloud formation in their atmospheres. For example, Figure \ref{fig:ijk} shows that L subdwarfs have much bluer optical and near-infrared (NIR) colours than L dwarfs. 

A three-dimension classification concept for ultracool dwarfs/subdwarfs was proposed by \citet{kirk05}. The core part of spectral types is spectral class and subtype which indicating $T_{\rm eff}$ and clouds. The prefix and suffix of spectral types are subclasses corresponding to different metallicity (e.g. d, sd, esd) and gravity (e.g. $\alpha, \beta, \gamma$), respectively. The metal subclass is more relevant to old ultracool subdwarfs which cover a wide range of metallicity and have similar gravity. The gravity subclass is more relevant to young ultracool dwarfs which cover a wider range of gravity but a smaller range of metallicity. 

The classification of L subdwarfs is mainly on their subtypes and metal subclasses. Subtypes of L subdwarfs are determined by comparing their red optical spectra to those of L dwarfs \citep{burg07,kirk10}. \citet{zha17a} classified L subdwarfs into ultra subdwarf (usdL), extreme subdwarf (esdL), and subdwarf (sdL) subclasses, corresponding to metallicity ranges of 
[Fe/H] $< -1.7$ (usdL); 
$-1.7 <$ [Fe/H] $< -1$ (esdL);
$-1 <$ [Fe/H] $< -0.3$ (sdL). 
Spectral features of each subclass are summarized in table 3 of \citet{zha17a}. Figure \ref{fig:ijk} shows that L subdwarfs of difference subclass are well separated by their optical to NIR colours. 

Figure \ref{fig:spec} shows optical to NIR spectra of usdL4, esdL4, sdL4, dL4, and red dL4 subclasses normalized in the optical. We can see that objects with lower metallicity have bluer spectra and larger suppression in the NIR. Note that the true relative flux between spectra of different subclass are contrary as shown in Figure \ref{fig:spec}. Figure 22 in \citet{zha18b} shows that L4 type spectra of different subclass have a similar $H$-band absolute magnitude. Therefore, a set of these L4 type spectra of different metal subclass normalized at $H$-band would show their true relative flux at a same distance  \citep[e.g. fig. 15 of][]{zha17a}. Although all classified as L4 type, the usdL4 have the highest flux at wavelengths below 1600 nm.  

The spectral subtypes of L subdwarfs are correlated to $T_{\rm eff}$ and mass within the same metal subclass, but it is not the case across different metal subclass. More metal-poor L subdwarfs tend to have higher $T_{\rm eff}$. L subdwarfs are hotter than similar-typed L dwarfs by around 100--400 K depending on different subclasses and subtypes \citep[e.g. fig. 4 of][]{zha18a}. 
An L4 type object with lower metallicity is more massive than L4 type objects with higher metallicities. For example, 2M1626 is a halo T-BD with a mass of $\sim$ 0.828 M$_{\odot}$ \citep{zha18a}. Meanwhile, a young early L type dwarf of an open cluster could fall into the planetary mass domain. 

The usdL and esdL subclasses as well as a small fraction of sdL subclass are kinematically associated with the Galactic halo. The sdL subclass are mostly associated with the thick disk.

\begin{figure*}
	\centering
	\includegraphics[width=\linewidth]{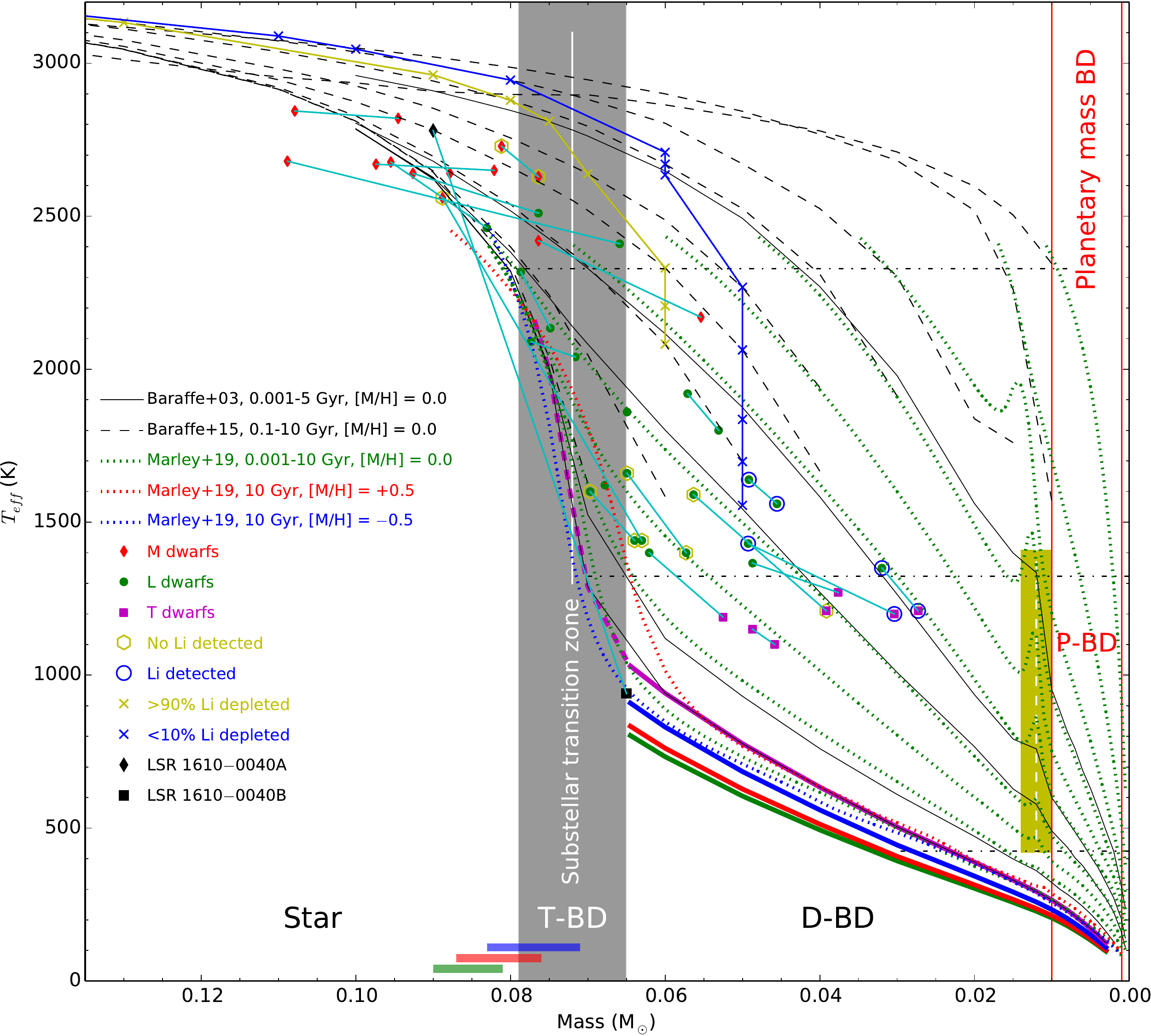}
	\caption{
Isochrones of $T_{\rm eff}$ of VLMS and substellar objects \citep[fig. 5;][]{zha19b}. The black solid lines are solar metallicity isochrones at 0.1, 0.5, 1, 5, and 10 Gyr \citep{bara03}. The black dashed lines are solar metallicity isochrones at 0.001, 0.01, 0.05, 0.1, 0.2, 0.3, 0.5, 1, and 5 Gyr  \citep{bara15}. The green dotted lines are solar metallicity isochrones at 0.001, 0.01, 0.04, 0.1, 0.2, 0.4, 1, 2, 4, and 10 Gyr (Marley et al. 2019, in prep.). The 10 Gyr isochrones with [M/H] = +0.5 and [M/H] = $-0.5$ (i.e. [Fe/H] $\approx -0.7$) from Marley et al. (2019) are overplotted and intersect with the 10 Gyr isochrone with [M/H] = 0.0 at 0.079 M$_{\odot}$, which defines the stellar-substellar boundary. A grey shaded band indicates the STZ (at $Z_{\odot}$) that hosts T-BDs, and have stars to its left and D-BDs to its right. A white line in the middle of the transition zone indicates the HBMM (0.072 M$_{\odot}$) predicted by \citet{chab97}. The red diamonds, the green circles, and the magenta squares are M7--T5 dwarfs that are in binary systems (joined with the cyan lines) thus have measurements of dynamical masses \citep{dupu17,lazo18}. Masses of an unresolved subdwarf binary, LSR 1610$-$0040AB (the black diamond and square), is from \citep{kore16}. Objects with and without lithium detection in their spectra are indicated with blue circles and yellow hexagons, respectively. The yellow and blue crosses indicate  $>90$ per cent and $<10$ per cent lithium depletion predicated by evolutionary models \citep{bara15}. The three dot-dashed lines indicate the average $T_{\rm eff}$ of  L0, T0, and Y0 dwarfs \citep{dupu13,dupu17}. The two vertical red lines indicate 1M$_{\rm Jup}$ and 0.01 M$_{\odot}$ (deuterium burning minimum mass). A yellow shaded band on the right indicates the deuterium burning transition zone (0.01--0.014 M$_{\odot}$), centred at 0.012 M$_{\odot}$ (the white dashed line). The 10 Gyr isochrones for T-BDs and D-BDs are highlighted with the magenta dashed and solid thick lines, respectively. The blue, red, and green solid thick lines indicate the 10 Gyr isochrones at 0.1, 0.01, and 0.001 $Z_{\odot}$, respectively. The blue, red, and green bars at the bottom indicate mass ranges of STZ at 0.1, 0.01, and 0.001 $Z_{\odot}$.
}
	\label{fig:isoc}
\end{figure*}

\section{Transitional and degenerate brown dwarfs}
The 10-Gyr isochrones of low-mass objects with different metallicity (Figure \ref{fig:burr01}) show that the $T_{\rm eff}$ \citep[and luminosity; figs 4 and 5 of][]{burr01} of VLMS increases with decreasing metallicity. However, it is reversed below a certain mass where isochrones with different metallicity intersect. The intersection mass is slightly higher at lower metallicity. A steep decline of $T_{\rm eff}$/luminosity as a function of mass is followed after the intersection point. This is because objects with mass below the intersection mass can not maintain steady hydrogen fusion for constant energy supply. 

A 10-Gyr isochorne of a certain metallicity intersects with the 10-Gyr isochrone of the metallicity just above/below its own at the mass just below/above the steady hydrogen burning minimum mass (SHBMM) of this certain metallicity. This defines the stellar/substellar boundary of different metallicity. The SHBMM is around 0.08, 0.083, 0.845, 0.855, and 0.875 M$_\odot$ at [Fe/H] = $-0.7, -1.3, -1.6, -1.8$, and $-2.3$, respectively \citep{zha17b}, according to evolutionary models \citep{bara97,chab97}.

The steep decline of $T_{\rm eff}$ as a function of mass on the 10-Gyr isochrones in Figure \ref{fig:burr01} stop at around 1000 K where the mass is around 0.066--0.083 M$_\odot$ \citep[for dashed line with base models of][]{alla95} corresponding to 1--0.001 Z$_\odot$, respectively. At below 1000 K, $T_{\rm eff}$ decreases slowly with decreasing mass. D-BDs are no massive enough to fuse hydrogen and rely on their initial thermal energy, would cool into the $\leq 1000$ K domain at 10 Gyr. Note that the $T_{\rm eff}$ of D-BDs also decreases with decreasing metallicity, because the lower opacity at lower metallicity leads to a faster dissipation of thermal energy.  

The `waterfall-like' feature in Figure \ref{fig:burr01} represents a narrow mass range between VLMS and D-BDs, where objects can not maintain their $T_{\rm eff}$/luminosity like stars and also do not cool as fast as D-BDs. Objects in this region are T-BDs \citep{zha18a}. T-BDs are fully convective and could reach temporary states of temperature and pressure sporadically for hydrogen fusion in their cores. As hydrogen is fused in low-rate thus such unsteady hydrogen fusion could lasts for $\gg$ 10 Gyr \citep[e.g. fig. 6;][]{burr11}, and slowly becomes the dominate energy sources to maintain their luminosity. Since the core temperature of a T-BD declines slowly over time. The efficiency of the fusion also declines very slowly over time. The efficiency of the fusion is very sensitive to the mass of T-BDs. Therefore, field T-BDs at a certain age could span in a wide $T_{\rm eff}$ range within a narrow mass range. Note the $T_{\rm eff}$ range of T-BDs is wider at older age or lower metallicity \citep{zha18b}. 

Fig. \ref{fig:teff} shows the 10 Gyr isochrones of $T_{\rm eff}$ of T-BDs and D-BDs with metallicity of [M/H] = +0.5, 0.0, and $-$0.5 predicted by the latest Sonora models (Marley et al. 2019, in prop.). The 10 Gyr isochrone with [M/H] = 0.0 intersects with the 10 Gyr isochrones with [M/H] = +0.5 and [M/H] = $-0.5$ at 0.079 M$_{\odot}$, which defines the stellar-substellar boundary at [M/H] = 0.0. Meanwhile, Fig. \ref{fig:gr} shows the 10 Gyr isochrones of gravity and radius of T-BDs and D-BDs with metallicity of [M/H] = +0.5, 0.0, and $-$0.5 (Marley et al. 2019, in prop.). The gravity maximum and radius minimum of dwarfs with [M/H]= 0.0 and 10 Gyr age are around 0.065 M$_{\odot}$ according to evolutionary models (Marley et al. 2019), which defines the boundary between T-BDs and D-BDs. Therefore, the mass range of T-BDs with solar metallicity is between 0.065 and 0.079 M$_{\odot}$ \citep{zha19b}. The substellar boundary is a mass range rather than a mass point. The HBMM predicted by \citet{chab97} around 0.072 M$_{\odot}$ is actually the middle of this mass range.

\section{The substellar transition zone}
The $T_{\rm eff}$ distribution of T-BDs is being stretched over time and forming a `waterfall-like' feature at around 0.07-0.09 M$_\odot$ (depending on metallicity; e.g. Figure \ref{fig:burr01}). The stretching is most significant among halo T-BDs which have $\sim$ 10 Gyr of evolutionary time. \citet{zha17b} plotted 10-Gyr iso-mass counters in a $T_{\rm eff}$ versus [Fe/H] space (Figure \ref{fig:tm}). We can see that for stars with the same mass, those have lower metallicity are also hotter. However, these iso-mass counters between those green lines are stretched towards the low temperature direction, and revealed a `substellar transition zone' - corresponding to the waterfall-like feature revealed by isochrones in Figure \ref{fig:burr01}. 

Halo T-BDs could be separated from VLMS by their $T_{\rm eff}$ and [Fe/H], as they have similar age around 10 Gyr. Halo T-BDs are in a narrow mass range, but they fall into a broad $T_{\rm eff}$ range. Halo T-BDs have $T_{\rm eff}$ between 2200--3000 and 1000 K, and spectral types of $\sim$ L3--T4. The $T_{\rm eff}$ of the stellar/substellar boundary is different across different metallicity (the left green line in Figure \ref{fig:tm}). SDSS J010448.46+153501.8 has the earliest spectral type (usdL1.5) among known halo T-BDs, it is also the most metal-poor and most massive substellar object know to date ([Fe/H] = $-$2.4; 0.086 M$_\odot$). 

Although halo T-BDs cover a wide range of spectral types, their number density in the Galaxy is extremely low, because of their narrow mass range on the initial mass function. For example, the transition zone at [Fe/H] = $-$1.7 is around 0.075--0.085 M$_\odot$ and covers a $T_{\rm eff}$ range of 2500--1000 K. Currently, there are ten known T-BDs discovered in the Galactic halo with spectral types from usdL1.5 to esdL7 \citep{zha18a,zha19a}. 
The majority of BDs are D-BDs. Those D-BDs in the halo have evolved into T5+ and Y subdwarfs.  

We draw an empirical stellar boundary on $i-J$ versus $J-K$ colour-colour plot (black dashed line in Figure \ref{fig:ijk}) according to observed colours of L subdwarfs separated by our stellar/substellar boundary in Figure \ref{fig:tm}. Figure \ref{fig:ijk} also shows that metal-poor substellar objects could be separated by optical and NIR colours sensitive to temperature and metallicity. 

Figure \ref{fig:isoc} shows solar metallicity isochrones for VLMS and BDs between 0.01 and 10 Gyr. We can see that objects with mass above $\sim$ 0.08 M$_{\odot}$ could maintain their $T_{\rm eff}$ over 10 Gyr. Objects in the grey band are stretched into a wide $T_{\rm eff}$ range. Meanwhile, objects with mass below $\sim$ 0.065 M$_{\odot}$ cool continuously. The different evolution of these objects were governed by their different energy supply sources. VLMS are above the SHBMM, therefore have constant energy supply from hydrogen fusion. T-BDs have long-lasting unsteady hydrogen fusion to partially replenish the dissipation of initial thermal energy. The efficiency and duration of the fusion could last for billions of years and is highly depending on their masses \citep{burr11}. D-BDs have no hydrogen fusion to replenish the dissipation of initial thermal energy. 

The existence of the substellar transition zone has impacts on observations and characterization of field BDs. It is difficult to distinguish field T-BDs by spectroscopy and atmospheric parameters, as their luminosity are dominated by their initial thermal energy and have mass/age degeneracy. L type field T-BD are mixed with L type D-BDs that are crossing the substellar transition zone. 

There is a lack of objects at the L/T transition \citep{burg07b}. This is firstly because the rapid evolution of BD atmospheres at $\sim$ 1200 K stretched the spectral subtype sampling. Secondly, the L/T transition is at the bottom of the substellar transition zone and next to the abundant D-BDs that crossed the transition zone \citep{zha18b}. 

L type D-BDs are younger than cooler T type D-BDs on average according to the isochrones in Figure \ref{fig:isoc}. However, if we compare the age of a field L dwarf sample and a field T dwarf sample. They should not show much difference. This because a large fraction of field L dwarf sample are T-BDs which are as old as field T dwarfs. 

The transition zones with incomplete lithium and deuterium fusions also exist among D-BD population, but have less impacts on properties of BDs than the hydrogen fusion transition zone \citep{zha19b}. The lithium transition zone is between 0.05-0.06 M$_{\odot}$ (Fig. \ref{fig:isoc}). The presents of 670.8-nm lithium absorption line is used to identify L type D-BDs with mass less than $\sim$ 0.05 M$_{\odot}$. The deuterium transition zone is between 0.01-0.014 M$_{\odot}$. D-BDs with mass below 0.01 M$_{\odot}$ ($\sim$ 10.5 M$_{\rm Jup}$) without any nuclear fusion could be referred to as planetary mass BDs (P-BDs). 


\section{Conclusions}
T-BDs are a population that different from both VLMS and D-BDs. They have unsteady hydrogen fusion to partially replenish the dissipation of their initial thermal energy, which led to a different evolution and impacted on observational properties of BD population. The substellar transition zone is a stretched temperature canyon of T-BDs in a narrow mass range due to unsteady hydrogen fusion. The substellar transition zone range from 2200--3000 K to 1000 K and from $\sim$ L3 to T4 types. The moderate evolution of T-BDs and the existence of the substellar transition zone need to be considered in the characterization of BD population. T-BDs with solar metallicity have mass between 0.065-0.079 M$_{\odot}$. D-BDs with solar metallicity have mass below 0.065 M$_{\odot}$. Metal-poor D-BDs have evolved to T5+ and Y subdwarfs with $T_{\rm feff} <$ 1000 K. The transition zones of incomplete lithium (0.05-0.06 M$_{\odot}$) and deuterium (0.01-0.014 M$_{\odot}$) fusions also exist in D-BD population.


\bibliographystyle{cs20proc}
\bibliography{zzhang_cs20.bib}

\end{document}